# Comments on the paper: Synthesis growth and characterization of copper mercury thiocyanate crystal [Indian J Pure & App Phys 49 (2011) 340-343]

Bikshandarkoil R. Srinivasan
Department of Chemistry, Goa University, Goa 403 206, INDIA
Email: srini@unigoa.ac.in

## Abstract

The unit cell parameters, infrared and UV-Vis spectral data reported in the paper by Vijayabhaskaran *et al* (Indian J Pure & App Phys 49 (2011) 340-343) cannot belong to the colourless crystalline compound formulated as copper mercury thiocyanate $CuHg(SCN)_4$ as claimed by the authors.

## Comment

Ammonium mercury thiocyanate $(NH_4)_2[Hg(SCN)_4]$ has long been used as a reagent to detect the presence of elements like Co, Ni, Cu, Zn and Cd with which it forms crystalline precipitates having characteristic shape and colour. The precipitates of general formula $AB(SCN)_4$ (A = Co, Ni, Cu, Zn, Mn; B = Hg) have been the subject of several research investigations over the past more than hundred years[1-16], since the first report on these compounds by Rosenheim and Cohen[1] appeared in 1901. This century long research has enhanced our understanding of the synthetic aspects, spectral characteristics, structural features and material applications of the $MHg(SCN)_4$ compounds. In these compounds which exhibit a three dimensional polymeric structure (Table 1), the thiocyanate anion functions as a bridge between a bivalent $3d$ metal ion namely M(II) and Hg(II) ion; the softer Hg(II) is linked to S sites of four symmetry related thiocyanates forming a $\{HgS_4\}$ tetrahedron and the bivalent $3d$ metal ion bonded to N sites of four different thiocyanate ligands resulting in a $\{MN_4\}$ tetrahedron excepting for Cu(II) and Ni(II). Bergman et al from the Bell Telephone Laboratories demonstrated in 1970 that the colourless $CdHg(SCN)_4$ and $ZnHg(SCN)_4$ exhibit useful nonlinear optical (NLO) properties[9].

Table 1 – Colour and geometry of metal in thiocyanate bridged $MHg(SCN)_4$ polymers

| Compound | Space group | Colour | Geometry of Hg(II) | Geometry of M(II) | Ref |
|---|---|---|---|---|---|
| $CoHg(SCN)_4$ | *I-4* | Blue | {$HgS_4$} tetrahedron | {$CoN_4$} tetrahedron | 7 |
| $MnHg(SCN)_4$ | *I-4* | Pale yellow | {$HgS_4$} tetrahedron | {$MnN_4$} tetrahedron | 14 |
| $ZnHg(SCN)_4$ | *I-4* | Colourless | {$HgS_4$} tetrahedron | {$ZnN_4$} tetrahedron | 13 |
| $CdHg(SCN)_4$ | *I-4* | Colourless | {$HgS_4$} tetrahedron | {$CdN_4$} tetrahedron | 12 |
| $CuHg(SCN)_4$ | *Pbcn* | Green | {$HgS_4$} tetrahedron | {$CuN_4$} square | 3 |
| $CuHg(SCN)_4$ | *C2/c* | Green | {$HgS_4$} tetrahedron | {$CuN_4$} square | 16 |
| $Cu_{0.16}Zn_{0.84}Hg(SCN)_4$ | -- | Purple | --- | {$CuN_4$} tetrahedron[a] | 5, 10 |
| $NiHg(SCN)_4 \cdot 2H_2O$ | -- | Blue | --- | {$NiN_4O_2$}[b] | 1, 5 |
| $NiHg(SCN)_4$ | -- | Yellow Green | --- | {$NiN_4S_2$}[b] | 5 |

Abbreviations: [a]Based on X-ray powder data which is very similar to $ZnHg(SCN)_4$; [b] the Ni(II) site is tetragonally distorted octahedron was proposed based on UV-Vis and magnetic data

Vijayabhaskaran et al decided to grow and investigate crystals of $CuHg(SCN)_4$ for NLO activity and reported[17] that by taking $CuCl_2$, $HgCl_2$ and KSCN in a 1:1:4 mole ratio in a mixture of water and ethanol, crystals of copper mercury thiocyanate, formulated as $CuHg(SCN)_4$ can be grown. An analysis of the paper shows that the results reported in this paper do not in any way concern with $CuHg(SCN)_4$ crystals and the paper is completely erroneous. To become convinced of this fact it is enough to first have a look at the photograph of the crystals in Fig. 1 in the reported paper. One can be very certain that the colorless crystals displayed in Fig. 1 can never be that of a cupric compound as it is well documented that copper(II) compounds are colored. It is to be noted that $CuHg(SCN)_4$ is a green coloured compound and two polymorphic modifications[3,16] of this compound have been structurally characterized. While the second polymorphic form (C2/c) of $CuHg(SCN)_4$ was reported very recently[16], the orthorhombic modification has been known for several years[3].

For formulating the crystals which they grew, as $CuHg(SCN)_4$ the authors used only unit cell parameters, IR and UV-Vis spectra. A comparison of the unit cell parameters given by the authors (Table 2), with those reported for the two different polymorphic forms[3,16] of $CuHg(SCN)_4$ clearly reveals that the unit cell parameters cannot be that for $CuHg(SCN)_4$.

Table 2 Comparison of Single Crystal X-ray data on CuHg(SCN)$_4$

| No | $a$ (Å) | $b$ (Å) | $c$ (Å) | $β$ (°) | V (Å$^3$) | Space Group | Ref |
|---|---|---|---|---|---|---|---|
| 1 | 9.03 | 7.68 | 15.15 | 90.0 | ---- | Pbcn | 3 |
| 2 | 9.0084(18) | 7.6993(6) | 15.1560(20) | 90.00(10) | 1051.2(4) | C2/c | 16 |
| 3 | **11.09** | **4.10** | **11.34** | **115.13** | **467** | Monoclinic | 17 |

**Note** - The data in **bold** are reported by Vijayabhaskaran *et al* in Ref. 17 where only the crystal system and not the space group is given. For details of why a beta value of 90.00 degree was chosen pl. read Ref. 16

That the crystals reported by Vijayabhaskaran *et al*[17] cannot be CuHg(SCN)$_4$, can be further confirmed from the infrared (IR) and UV-Vis spectral data which in no way resemble the known spectral characteristics of the two known modifications of CuHg(SCN)$_4$ given in literature[4-6,11,16]. Since both modifications of CuHg(SCN)$_4$ are known to crystallize in centrosymmetric space groups, the observed SHG property in the form of 532 nm green emission on excitation with 1064 nm laser beam conclusively indicates that the crystals grown by the authors cannot be CuHg(SCN)$_4$.

The questionable nature of the results presented in the title paper is evident from the experimental details of crystal growth, which does not provide quantities (in terms of weight) of reactants, volume of solvent used namely water and ethanol and yield of final product but instead gives an equation (see below eq. 1) for the formation of CuHg(SCN)$_4$. Immediately thereafter the authors write that 'The solution was filtered twice to remove any insoluble impurities. Then the purity of the compound was increased by successive recrystallization processes'.

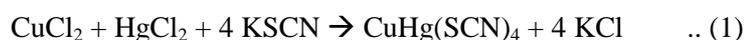

$$CuCl_2 + HgCl_2 + 4\ KSCN \rightarrow CuHg(SCN)_4 + 4\ KCl \qquad .. (1)$$

Since CuHg(SCN)$_4$ is the only insoluble material[8] in the above equation, it is not clear from the experimental details in the paper as to how the authors increased the purity of their crystalline product. It is to be noted that the reported synthesis[1,2] of CuHg(SCN)$_4$ employs a procedure wherein an aqueous solution of a copper(II) salt is added into an aqueous solution containing [Hg(SCN)$_4$]$^{2-}$ (bound thiocyanate) to obtain the green CuHg(SCN)$_4$ in quantitative yield. The same procedure is also recommended for the gravimetric estimation[8] of Cu(II) as CuHg(SCN)$_4$. The reason for following such a protocol is due to the well known reactive nature of aqueous solutions of Cu(II) with free thiocyanate ions[18,19] (see equation 2) resulting in the formation of a black precipitate of [Cu(NCS)$_2$], which on standing slowly changes to the insoluble colorless Cu(SCN) with the formation of thiocyanogen (SCN)$_2$. This chemistry aspect has probably not been taken into due consideration by Vijayabhaskaran *et al*[17].

$$Cu^{2+} (aq) + 2 (SCN)^- (aq) \rightarrow Cu(NCS)_2$$

$$\rightarrow Cu(SCN) + \tfrac{1}{2} (SCN)_2 \quad .. (2)$$

The black insoluble [Cu(NCS)$_2$] can probably be one way in which the Cu(II) content was lost by the authors (as an impurity) since all the reagents employed in the synthesis are known to be water soluble. Another possible explanation for a colorless crystalline product obtained by authors can be that they filtered off the insoluble Cu-containing product thinking it as impurity. Although the exact nature of the final product cannot be commented upon in the absence of details of product yield, the colorless nature of the crystals clearly confirms the absence of Cu(II) in the product.

That the authors have not taken into account the above chemistry of Cu(II) with free thiocyanate ions can be evidenced from a synthesis of CuCo(SCN)$_4$ reported by two of these authors in another paper[20] according to equation 3. Here again it is noted that the authors filtered twice to remove insoluble impurities.

$$CuCl_2 + CoCl_2 + 4\ KSCN \rightarrow CuCo(SCN)_4 + 4\ KCl \quad .. (3)$$

It is to be noted that bivalent metals of the 3*d* series (Mn(II), Fe(II), Co(II), Ni(II), Cu(II) and Zn(II)) are all known to bind to (SCN)$^-$ at the N-site and not at the sulfur end which is linked to a softer metal like Hg. The reported claim of attaining a polymeric AB(SCN)$_4$ structure type, by using two A type metal ions namely Cu(II) and Co(II) for synthesis of CuCo(SCN)$_4$[20] and Zn(II) and Mn(II) for synthesis of ZnMn(SCN)$_4$[21] is not only meaningless but just impossible, because both pairs of metal ions are known to bind at the N-site of (NCS)$^-$ forming isothiocyanates.

In order to verify the reported claim of a synthesis of CuCo(SCN)$_4$, the reaction of an aqueous solution containing equimolar mixtures of Cu(II) and Co(II) ions was investigated with four moles of thiocyanate. This resulted in the immediate precipitation of all the available Cu(II) as [Cu(NCS)$_2$] according to equation 2 leaving behind only Co(II) in solution. The loss of copper clearly indicates that the synthetic scheme (eq.3) given by authors in their work is incorrect and hence the claim of the authors of the synthesis of a bimetallic thiocyanate compound namely $CuCo(SCN)_4$ is totally erroneous[20]. However, the product of the reaction which does not contain any Cu was not only wrongly formulated but claimed as a new non-linear-optical material based on its unit cell parameters discussed under the title single crystal XRD. It is inappropriate to characterize a new compound based on just cell parameters. In a recent paper Fleck and Petrosyan[22] have given several examples of incorrectly characterized compounds based on unit cell parameters. In the case of $CuCo(SCN)_4$ the chemistry of the crystal synthesis itself is wrong.

The synthesis of $ZnMn(SCN)_4$ reported by the same group[21] by taking reactants in a 1:1:1 mole ratio according to equation 4 can also be very easily disproved.

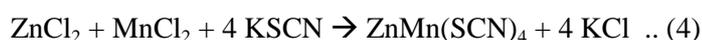

$ZnCl_2 + MnCl_2 + 4\ KSCN \rightarrow ZnMn(SCN)_4 + 4\ KCl$ .. (4)

Note the disparity in the experimental procedure for reactants (1:1:1) and the reaction scheme. A comparison of the unit cell parameters (cell volume V = 1256 Å$^3$) reported by Paramasivam *et al*[21] with those of the reported single crystal data for $ZnHg(SCN)_4$ (V= 546.36(6)Å$^3$) and $MnHg(SCN)_4$ (V=547.5(3)Å$^3$) reveals the dubious nature of the work (Table 3). The replacement of a larger ion like Hg(II) by Mn(II) can never result in doubling of the unit cell volume indicating that data in ref. 21 are dubious. The X-ray powder pattern of the alleged $ZnMn(SCN)_4$ given by the authors of Ref. 21 does not compare with the pattern for tetragonal $AB(SCN)_4$ reported by Wang et al[15].

Table 3 Comparison of unit cell data of $AB(SCN)_4$ compounds crystallizing in I-4 space group

| No | Compound | *a* (Å) | *b* (Å) | *c* (Å) | V (Å$^3$) | Ref |
|---|---|---|---|---|---|---|
| 1 | $CoHg(SCN)_4$ | 11.109 | 11.109 | 4.379 | 540.41 | 7 |
| 2 | $MnHg(SCN)_4$ | 11.324(3) | 11.324(3) | 4.270(2) | 547.5(3) | 14 |
| 3 | $ZnHg(SCN)_4$ | 11.0912(4) | 11.0912(4) | 4.4414(4) | 546.36(6) | 13 |
| 4 | $CdHg(SCN)_4$ | 11.487(3) | 11.487(3) | 4.218(1) | 556.6(2) | 12 |
| **5** | **$ZnMn(SCN)_4$** | **12.0835** | **12.131** | **8.569** | **1256** | **21** |

**Note** - The data in **bold** (which are very odd) are reported by Paramasivam *et al* in Ref. 21

*For entry Nos. 1 to 4 the structures are solved and the CIF files are available. For entry No. 5 the paper claims it is single crystal XRD work. However no data other than unit cell parameters are provided. Only the crystal system (and no space group) is mentioned as tetragonal. No CIF file is available for this work.*

The presence or absence of water in a transition metal thiocyanate can make a lot of difference as can be seen in the case of Ni(II) where the anhydrous compound $NiHg(SCN)_4$ is yellow-green in colour while the dihydrate $NiHg(SCN)_4 \cdot 2H_2O$ is blue in colour (Table 1) and thus exhibit different visible spectra[5]. A scrutiny of the IR spectrum of each of the compounds described in Ref. 17, 20, 21 gives the first indication that the proposed formula of the compound is incorrect. Each one of these compounds show an intense signal in its IR spectrum in the O-H region which is never to be expected in a $AB(SCN)_4$ compound which has no oxygen atom in its formula. Another unusual feature is that the UV-Vis-NIR spectra of the alleged compounds $CuHg(SCN)_4$[17] $CuCo(SCN)_4$[20] and $ZnMn(SCN)_4$[21] are nearly identical.

It is not clear as to why the authors chose to grow crystals of a well-known centrosymmetric compound like $CuHg(SCN)_4$ for investigating NLO property, the green colour of which makes it not very suitable for the intended study namely observation of SHG in the form of a green signal. It is quite unfortunate to note that in all the three examples discussed in this comment the authors give a chemical equation for synthesis of the compound without taking into account the chemistry of the reactants in the equation, and consequently the experimental evidence for the compounds described is not in accordance with the proposed formula.